\documentclass[review]{elsarticle}

\usepackage{lineno,hyperref}

\usepackage{amsmath}
\usepackage{physics}
\usepackage{booktabs}

\modulolinenumbers[5]

\journal{Journal of Luminescence}

\begin{document}

\begin{frontmatter}

\title{Hyperfine interaction and coherence time of praseodymium ions at the site 2 in yttrium orthosilicate}


\author[mymainaddress,mysecondaryaddress]{Yi-Xin Xiao}

\author[mymainaddress,mysecondaryaddress]{Zong-Quan Zhou\corref{mycorrespondingauthor1}}
\cortext[mycorrespondingauthor1]{Corresponding author}
\ead{zq\_zhou@ustc.edu.cn}

\author[mymainaddress,mysecondaryaddress]{Yu Ma}

\author[mymainaddress,mysecondaryaddress]{Tian-Shu Yang}

\author[mymainaddress,mysecondaryaddress]{You-Zhi Ma}

\author[mymainaddress,mysecondaryaddress]{Chuan-Feng Li\corref{mycorrespondingauthor2}}
\cortext[mycorrespondingauthor2]{Corresponding author}
\ead{cfli@ustc.edu.cn}

\author[mymainaddress,mysecondaryaddress]{Guang-Can Guo}

\address[mymainaddress]{CAS Key Laboratory of Quantum Information, University of Science and Technology of China, Hefei, 230026, P.R.China\\}
\address[mysecondaryaddress]{CAS Center For Excellence in Quantum Information and Quantum Physics, University of Science and Technology of China, Hefei, 230026, P.R.China\\}

\begin{abstract}
Praseodymium (Pr$^{3+}$) ions doped in the site 1 of yttrium orthosilicate (Y$_2$SiO$_5$) has been widely employed as the photonic quantum memory due to their excellent optical coherence and spin coherence. While praseodymium ions occupying the site 2 in Y$_2$SiO$_5$ crystal have better optical coherence as compared with those at site 1, which may enable better performance in quantum memory. Here we experimentally characterize the hyperfine interactions of the ground state $^3$H$_4$ and excited state $^1$D$_2$ of Pr$^{3+}$ at site 2 in Y$_2$SiO$_5$ using Raman heterodyne detected nuclear magnetic resonance. The Hamiltonians for the hyperfine interaction are reconstructed for both ground state $^3$H$_4$ and excited state $^1$D$_2$ based on the Raman heterodyne spectra in 201 magnetic fields. The two-pulse spin-echo coherence lifetime for the ground state is measured to be 2.6$\pm$0.1 ms at site 2 with zero magnetic field, which is more than five times longer than that at site 1. The magnetic fields with zero first order Zeeman shift in the hyperfine transition for Pr$^{3+}$ at site 2 in Y$_2$SiO$_5$ are identified.
\end{abstract}


\end{frontmatter}

\linenumbers

\section{\label{intro}Introduction}

Optical quantum memory is a core component of quantum repeaters \cite{Briegel-Repeaters-1998} to realize large-scale quantum network, which overcomes the photon loss from imperfect transmission channels \cite{Sangouard-repeaters-2011}. Besides, optical quantum memory can be instrumental for the development of many devices in quantum computing \cite{Shor-decoherence-1995} and multiple photon sources \cite{Hong-one-photon-1986}.

Among various candidate physical systems for quantum memory, rare-earth-ion-doped crystals (REICs) are promising material for quantum memory due to long coherence time \cite{Roger-resolution-2002}, long hyperfine population lifetimes \cite{Nilsson-Hole-burning-2004} and easy integration \cite{Seri_Heralded_2019,Liu_waveguide_2020}. There is a large inhomogeneous broadening of the optical transitions in REICs, which is ideal for multiplexed storage in temporal and spectral domain \cite{Afzelius-Multimode-2009,Sinclair-Multiplexing-2014,tang_storage_2015}.

Pr$^{3+}$:Y$_2$SiO$_5$ is an attractive material for solid-state quantum memory because it offers three hyperfine ground levels for on-demand spin-wave storage at zero magnetic field. This material has been employed for the demonstrations of spin-wave quantum storage of photonic time-bin qubits \cite{gundogan_solid_2015}, entanglement between a photon and a collective spin excitation in a rare-earth-ion-doped ensemble \cite{Kutluer-Entanglement-2019}, multiple degree-of-freedom quantum memory \cite{yang_multiplexed_2018}, long storage time for classical light \cite{Heinze-stopped-2013}, and quantum correlations between a cold atomic ensemble and a REIC \cite{maring_photonic_2017}. The reasons for the wide use of this crystal are long lifetimes of these hyperfine levels \cite{holliday_spectral_1993}, narrow homogeneous width of the optical transition and small inhomogeneous broadening for the hyperfine transitions in the ground state \cite{equall_homogeneous_1995}.

As a monoclinic crystal, Y$_2$SiO$_5$ has a structure given by the $C2/c$ space group which has eight formula units of Y$_2$SiO$_5$ per unit cell \cite{noauthor_crystal_1999}. In consequence, there are eight different locations where praseodymium could be substituted for yttrium. These locations can be divided to two sites, i.e. site 1 and site 2, which can be directly identified by its optical transition wavelength \cite{equall_homogeneous_1995}. Each site can be further divided into two sets which are not equivalent in a magnetic field and related to each other by a two-fold rotation about the C$_2$ axis. Most previous experiments are performed with Pr$^{3+}$ at site 1 which has the stronger oscillator strength. The Pr$^{3+}$ ion at site 2 in Y$_2$SiO$_5$ crystal has a narrower optical homogeneous linewidth than that at site 1 with both zero field and a weak magnetic field \cite{Nilsson-Hole-burning-2004,equall_homogeneous_1995}. Therefore, site-2 Pr$^{3+}$ ions in Y$_2$SiO$_5$ may provide better performance in optical storage compared with those at site 1. To further utilize the zero first order Zeeman hyperfine transitions \cite{fraval_method_2004,longdell_characterization_2006} for long-term optical storage, the hyperfine interaction should be characterized for searching of such critical fields. However, the details about the hyperfine interaction and spin coherence are absent so far.

In this work, we present the characterization of the hyperfine interaction for the ground state $^3$H$_4$ and excited state $^1$D$_2$ of Pr$^{3+}$ at site 2 in Y$_2$SiO$_5$ by Raman heterodyne detected nuclear magnetic resonance (NMR) \cite{longdell_hyperfine_2002,mlynek_raman_1983,wong_raman_1983,ma_raman_2018,Ligand-Ahlefeldt-2009,mcauslan_reducing_2012}. Besides, we measure the spin coherence time of the ground-state hyperfine transition of Pr$^{3+}$ at site 2 in Y$_2$SiO$_5$ in zero magnetic field and search critical magnetic fields that can extend the spin coherence lifetime.

\section{\label{theo}Theory of hyperfine interactions}

\begin{figure}[tb]
\includegraphics[width=\textwidth]{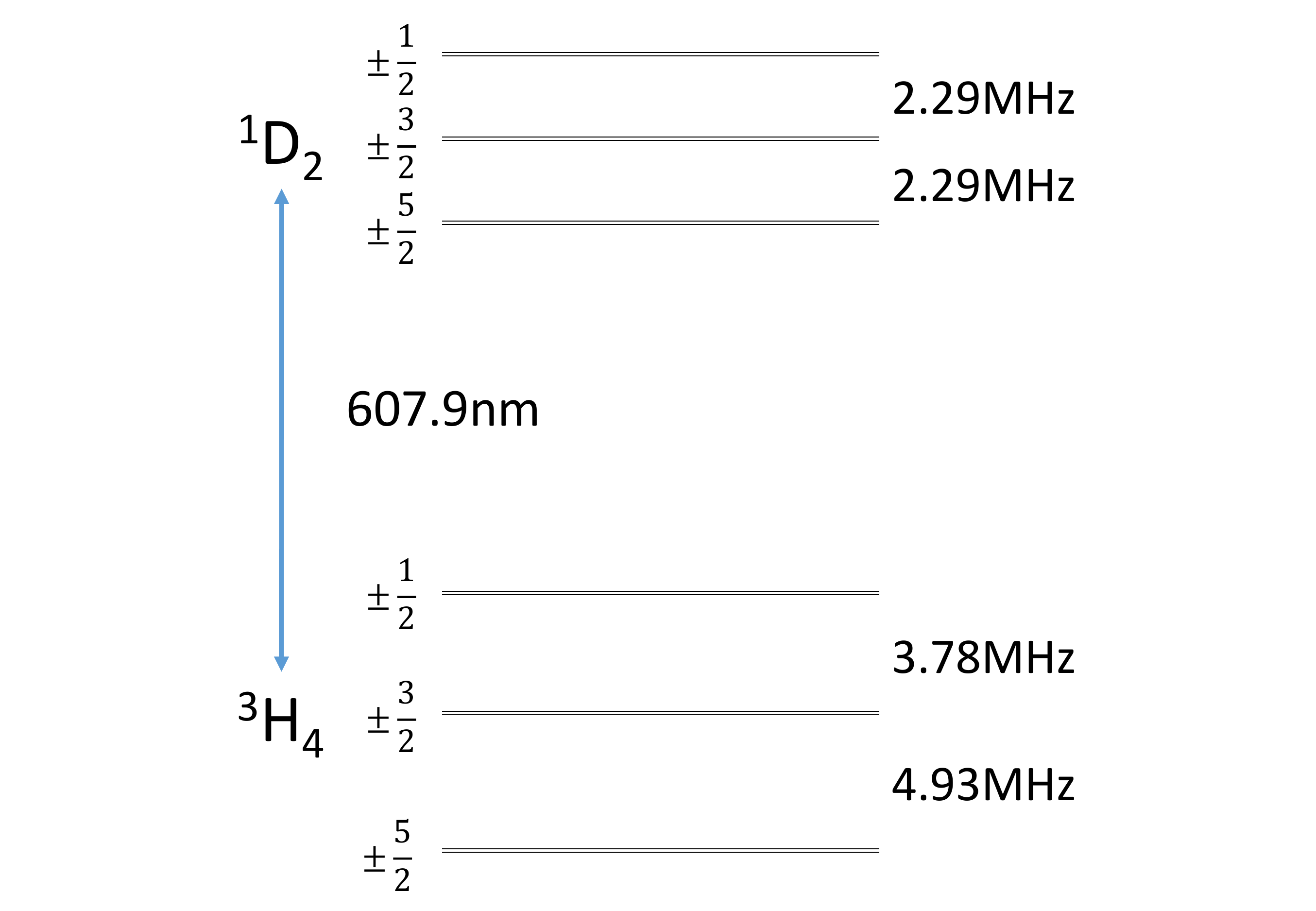}
\caption{\label{fig:level}The hyperfine structure of the ground state $^3$H$_4$ and excited state $^1$D$_2$ of Pr$^{3+}$ at site 2 in Y$_2$SiO$_5$.}
\end{figure}

Here we briefly review the theory of hyperfine interactions of REICs, which has been detailed in many previous publications  \cite{longdell_characterization_2006,longdell_hyperfine_2002,ma_raman_2018}. The Hamiltonian for nucleus and f electrons of rare-earth ions has the form
\begin{equation}
H = H_{FI} + H_{CF} + H_{HF} + H_{Q} + H_{z} + H_{Z}.\label{eq:original H}
\end{equation}
These six terms represent the free ion, crystal field, hyperfine, nuclear quadrupole, electronic Zeeman and nuclear Zeeman effect, respectively. The free ion and crystal field terms which determine the electronic energy levels play a major role on the Hamiltonian. The others could be regarded as a perturbation which causes the hyperfine structure for the electronic energy levels. As Pr$^{3+}$ is a non-Kramers ion, the degeneracy of multiplets is broken and each component of the multiplets can be considered as an orbital singlet. The last four terms in Eq. \ref{eq:original H} are small enough to be a perturbation that engenders the appearance of the hyperfine and magnetic effects. Since the hyperfine and electronic Zeeman terms only contribute to the effective spin Hamiltonian in the second order, using second order perturbation theory enables us to obtain the effective spin Hamiltonian as follows \cite{longdell_characterization_2006,longdell_hyperfine_2002,cruzeiro_characterization_2018}
\begin{equation}
\begin{split}
H &= B\cdot Z\cdot B + B\cdot M\cdot I + I\cdot Q\cdot I\\
&\approx B\cdot M\cdot I + I\cdot Q\cdot I, \label{eq:effective H}    
\end{split}
\end{equation}
where $B$ is the external magnetic field, $I$ is the nuclear spin operator, $M$ and $Q$ are the effective Zeeman and quadrupole tensors. The term $B\cdot Z\cdot B$ is the quadratic Zeeman term and discounted since it has no effect on the hyperfine splittings. The detailed parameterization is presented in Sec. \ref{fitting}.

Pr$^{3+}$ occupies two sites in the Y$_2$SiO$_5$, which have different absorption centered at 605.977 nm (site 1) and 607.934 nm (site 2) from $^3$H$_4$ to $^1$D$_2$ respectively \cite{equall_homogeneous_1995}. The zero magnetic field energy level diagram at site 2 is illustrated in Fig. \ref{fig:level}. 
Raman heterodyne signals can be accessed at 2.29, 3.78, 4.93 MHz in zero magnetic field. Specially, $\ket{\pm1/2}_e\longleftrightarrow\ket{\pm3/2}_e$ and $\ket{\pm3/2}_e\longleftrightarrow\ket{\pm5/2}_e$ have the same hyperfine energy level spacings, which are difficult to distinguish in Raman heterodyne signals. With the application of an external magnetic field, each degenerate pair of the hyperfine energy levels splits.

\section{\label{exp}Experiment}

\begin{figure}[tb]
\includegraphics[width=\textwidth]{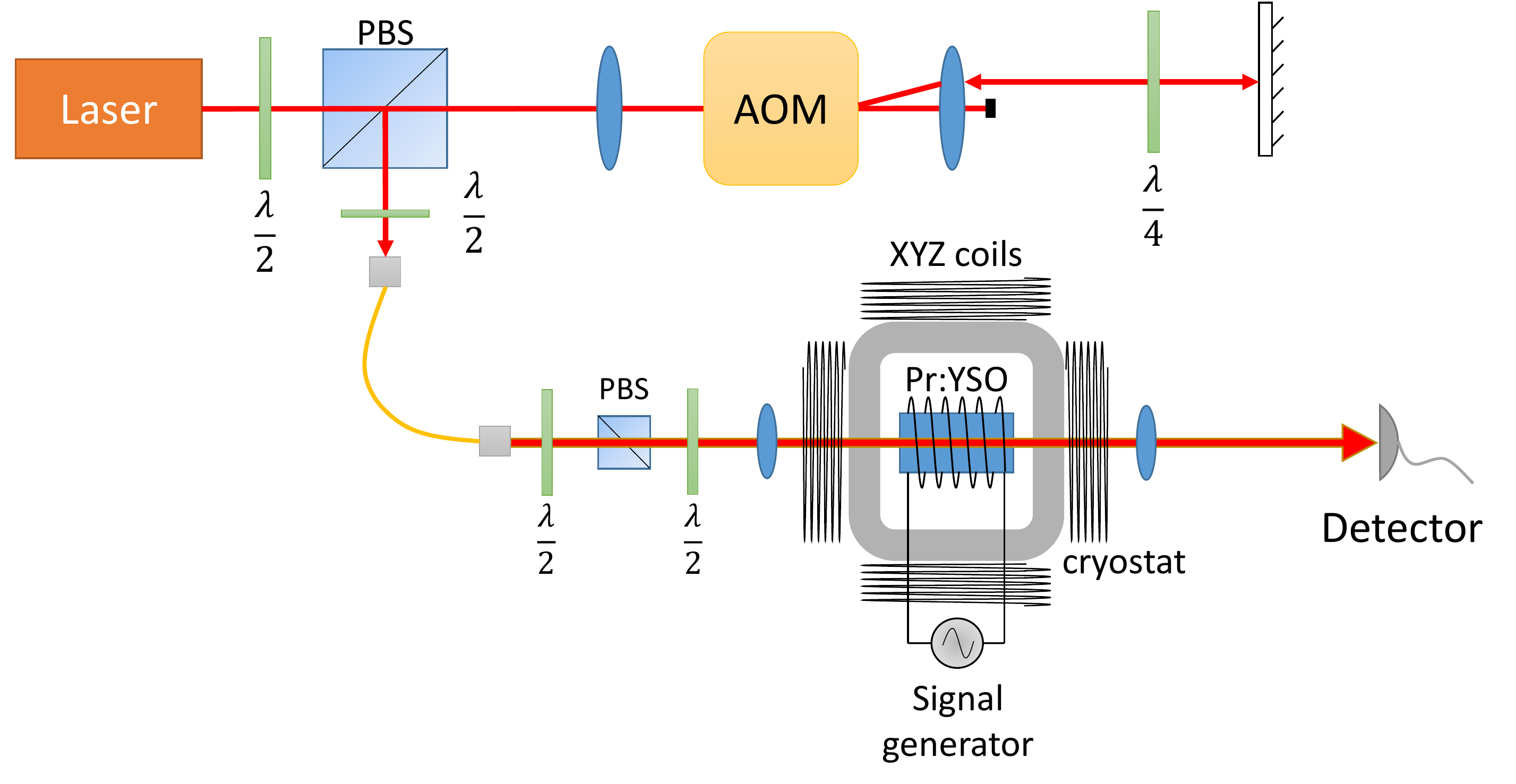}
\caption{\label{fig:setup}The setup of Raman heterodyne detected NMR experiment of Pr$^{3+}$ at site 2 in Y$_2$SiO$_5$. The input laser penetrates into the crystal inside a cryostat surrounded by XYZ coils. The frequency of the laser is controlled by an acousto-optic modulator (AOM) in a double-pass conﬁguration. The polarization of the beam in front of the cryostat is controlled by the half wave plate ($\lambda/2$) with the purpose of the highest absorption. A RF field is generated by a 6-turn coil wrapped around the crystal, which aims at driving the hyperfine transitions. The signal light generated from the crystal is captured by the photodetector.}
\end{figure}

The experimental setup is showed in Fig. \ref{fig:setup}. The Pr$^{3+}$ doped Y$_2$SiO$_5$ crystal (doping level: 0.05\%) was cut along the crystal axes D$_1$, D$_2$ and b with the sizes of $2.5\times6\times8$ mm$^3$. The crystal was placed in a cryostat for spectrum measurements and cooled to approximately 4.3 K.

To obtain Raman heterodyne spectra, a laser resonant with the optical transition at frequency $f_L$ was incident on the sample. A RF field that drove the hyperfine transition at frequency $f_{RF}$ was delivered by a six-turn coil wrapped around the sample. A Raman scattered light at frequency $f_L\pm f_{RF}$ was emitted, which propagated together with the transmitted light at frequency $f_L$ \cite{mlynek_raman_1983,wong_raman_1983}.

The light came from a frequency doubled semiconductor laser with a linewidth of less than 10 kHz. The input beam had a power of 9 mW in front of the cryostat. The beams that propagated along the $C_2$ axis (identical to the b axis) of the sample were then coupled into a single mode fiber and detected by a photodetector. The Raman heterodyne signal was generated from a beat between the transmitted light and the scattered light and then obtained by demodulation.

\begin{figure}[tb]
\includegraphics[width=\textwidth]{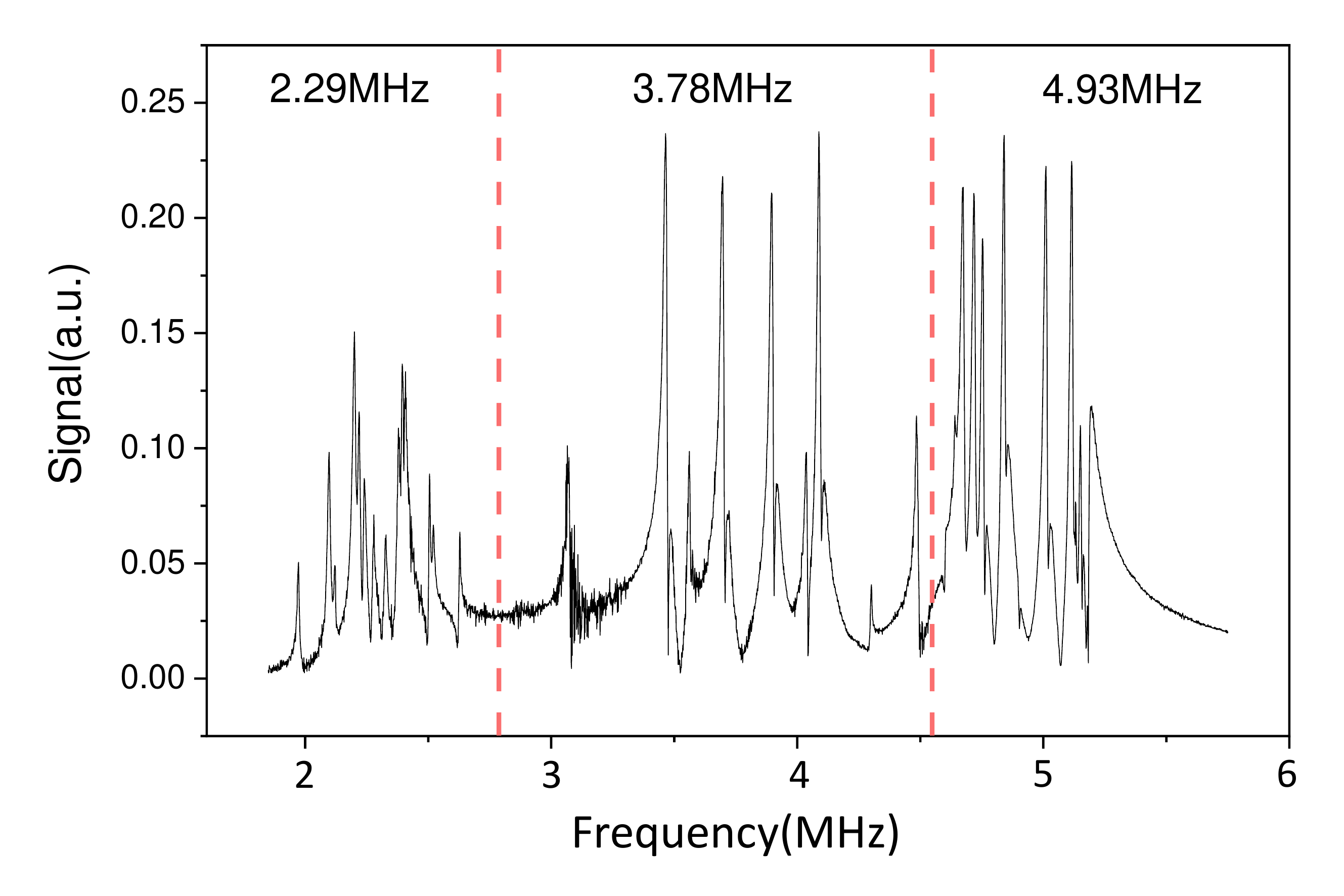}
\caption{\label{fig:raman}The typical Raman heterodyne spectrum for both ground and excited states with the external magnetic field $B=[-40,-13,68]G$. 8 peaks can be clearly observed in the ground state at 3.78 MHz or 4.93 MHz, whereas there are 16 peaks for the excited state at 2.29 MHz. These three parts are divided by the red dash lines.}
\end{figure}

The cryostat was located in XYZ coils which generated three-dimensional magnetic fields of 80 Gauss in order to split the degenerate hyperfine energy levels. The D$_1$, D$_2$ and b axes are close to the x, y and z axes of the magnet respectively. Similar to the experiment for site 1 \cite{longdell_hyperfine_2002}, the magnetic fields were rotated in a spiral form 
\begin{equation}
B = 
\left[
\begin{matrix}
B_0\sqrt{1-t^2}\cos(6\pi t)\\
B_0\sqrt{1-t^2}\sin(6\pi t)\\
B_0t
\end{matrix}
\right], t\in[-1,1].
\end{equation}
There were 201 equidistant points for t $\in$ [-1,1]. Because 3.78 MHz and 4.93 MHz of the two ground-state transitions and 2.29 MHz of the excited-state transition were close enough in the frequency domain, all hyperfine transitions were obtained in one spectrum [See Fig. \ref{fig:raman}]. When there was a magnetic field applied, a degenerate hyperfine energy level split to a pair due to the Zeeman effect, as a result each NMR peak split into four. However, there were eight peaks observed in the experiment because of the two different orientations of ions in the crystal. It was special that 16 NMR peaks were detected at around 2.29 MHz, which were a mixture of splittings in $\ket{\pm1/2}_e\longleftrightarrow\ket{\pm3/2}_e$ and $\ket{\pm3/2}_e\longleftrightarrow\ket{\pm5/2}_e$ transitions.

\section{\label{fitting}Fitting procedure}

\begin{figure*}[tb]
\includegraphics[width=\textwidth]{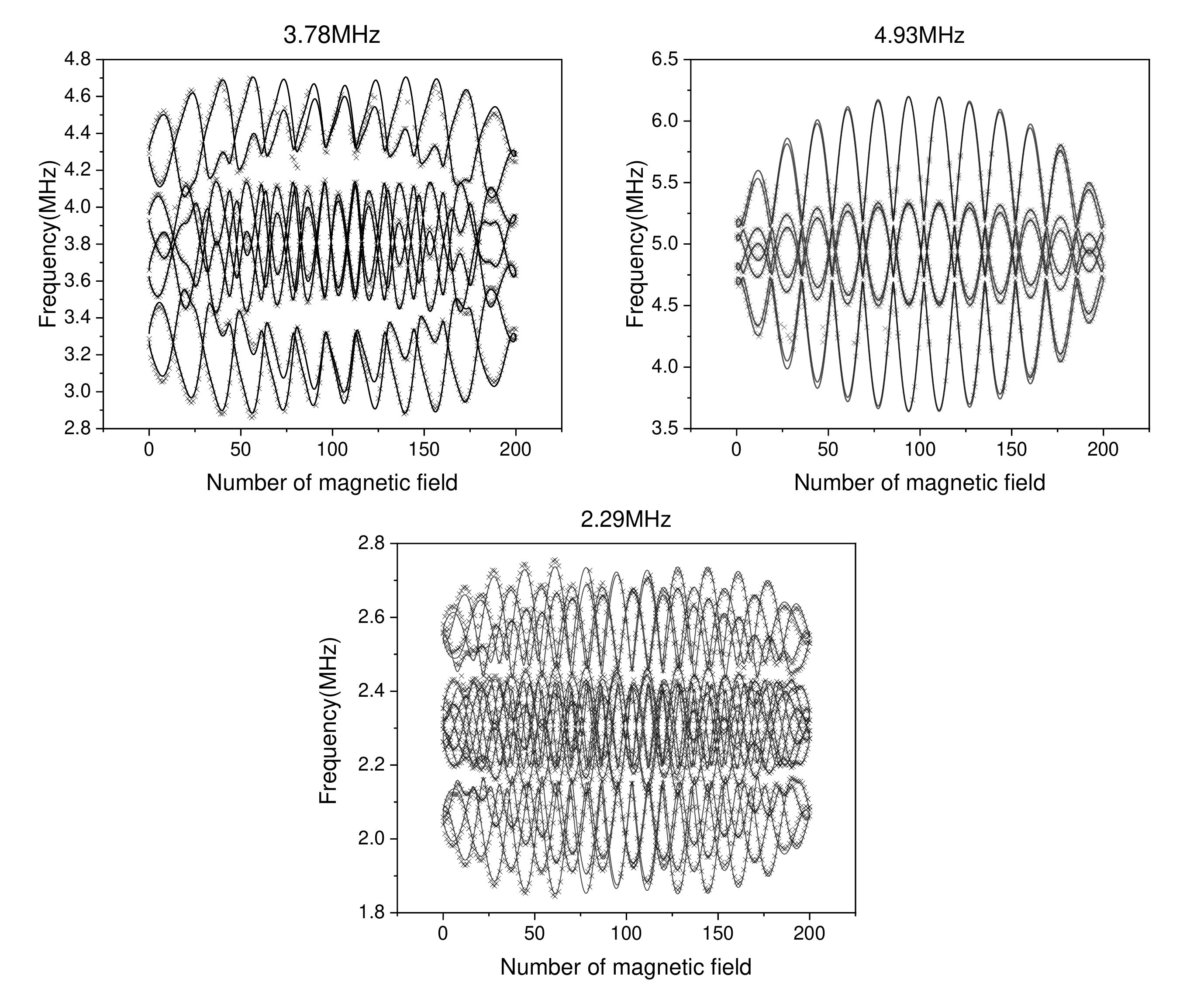}
\caption{\label{fig:spectral}The peak positions from the Raman heterodyne spectra for the ground and excited state of site-2 Pr$^{3+}$ in Y$_2$SiO$_5$ at 201 magnetic fields. The solid lines are the peak positions calculated from the fitting results, and the points are detected peak positions.}
\end{figure*}

In Fig. \ref{fig:spectral}, the positions of the corresponding peaks are selected from the recorded spectra for the fitting procedure. Using simulated annealing method \cite{longdell_hyperfine_2002,kirkpatrick_optimization_1983} to obtain the spin Hamiltonian can ensure that the fitting results conforms to the global solution. The Hamiltonian should be parametrized before the fitting procedure, which is the same as that for site 1 \cite{longdell_hyperfine_2002}.

Two parameters are required to determine the $Q$ tensor in the principal axis coordinate system, as the following parametrization:
\begin{equation}
Q_1 = R({\alpha _Q},{\beta _Q},{\gamma _Q})\left[
\begin{matrix}
{-E}&0&0\\
0&E&0\\
0&0&D
\end{matrix}
\right]
R^T({\alpha _Q},{\beta _Q},{\gamma _Q}),
\end{equation}
where $R({\alpha},{\beta},{\gamma})$ is the rotation matrix which is defined by the three Euler angles $(\alpha,\beta,\gamma)$ for ZYZ convention, 
\begin{equation}
\begin{split}
R({\alpha},{\beta},{\gamma}) =& \left[
\begin{matrix}
\cos\alpha&{-\sin\alpha}&0\\
\sin\alpha&\cos\alpha&0\\
0&0&1
\end{matrix}\right]\times
\left[
\begin{matrix}
\cos\beta&0&\sin\beta\\
0&1&0\\
{-\sin\beta}&0&\cos\beta
\end{matrix}\right] \\
&\times\left[
\begin{matrix}
\cos\gamma&{-\sin\gamma}&0\\
\sin\gamma&\cos\gamma&0\\
0&0&1
\end{matrix}\right].
\end{split}
\end{equation}
The Zeeman tensor has the same form as the $Q$ tensor:
\begin{equation}
M_1 = R({\alpha _M},{\beta _M},{\gamma _M})\left[
\begin{matrix}
g_1&0&0\\
0&g_2&0\\
0&0&g_3
\end{matrix}
\right]
R^T({\alpha _M},{\beta _M},{\gamma _M}),    
\end{equation}
where $(g_1,g_2,g_3)$ are Land$\Acute{e}$ g-factors and $({\alpha _M},{\beta _M},{\gamma _M})$ are another Euler angles for $M$ tensor. 

Because of the $C_2$ symmetry, the $Q_2$ and $M_2$ tensors for another set of ions are related to the $Q_1$ and $M_1$ tensors by rotating $\pi$ about the $C_2$ axis. There is a misalignment between the direction of the $C_2$ axis and the z axis defined by the XYZ coils. The $Q_2$ and $M_2$ tensors can be obtained by the $\pi$-rotation on $Q_1$ and $M_1$ tensors \cite{Lovri_spin_2012}: 
\begin{equation}
\begin{split}
M_2 &= R_{C_2}\cdot M_1\cdot R_{C_2}^T,\\
Q_2 &= R_{C_2}\cdot Q_1\cdot R_{C_2}^T,\\
R_{C_2} &= R(\phi,\theta,0)\cdot R(\pi,0,0)\cdot R^T(\phi,\theta,0),\\
\end{split}
\end{equation}
where $\theta$ represents the angle between the $C_2$ axis and the $z$ axis and $\phi$ is the angle between the projection of the $C_2$ axis onto the $xy$ plane and the $x$ axis. In consequence, two sets of ions have the Hamiltonians
\begin{equation}
\begin{split}
&H_1 = B\cdot M_1\cdot I + I\cdot Q_1\cdot I,\\
&H_2 = B\cdot M_2\cdot I + I\cdot Q_2\cdot I.
\end{split}\label{eq:final H}
\end{equation}

In the fitting procedure, the initial parameters are firstly set as random values which are roughly close to those of site 1 \cite{longdell_hyperfine_2002}. Next, the parameters are optimized to minimize the error between the calculated and experimental spectra using simulated annealing algorithm. 

It is worth noting that the Hamiltonian parameters of the excited state are fitted by 16 peaks from the two C$_2$-symmetric subsites due to the identical level splittings. The order of excited-state energy levels is unclear and it is impossible to determine which zero-field energy level splitting and subsite each transition corresponds to in the current experiment. Therefore the signs of the diagonal elements of $Q$ tensor in the excited state cannot be determined. Raman heterodyne detection cannot ensure that the excited-state Hamiltonians are related to the ground-state Hamiltonians in the same subsite \cite{ma_raman_2018}. The Raman heterodyne spectra are constant under certain transformations of the Hamiltonian, which are derived from the symmetry of the Hamiltonian \cite{cruzeiro_characterization_2018}. There are some remaining ambiguities in the spin Hamiltonian. First, the hyperfine Hamiltonians for the excited state and the ground state may not correspond to the same subsites due to two C$_2$-symmetric subsites. Spectral-hole burning measurements can help to eliminate this ambiguity \cite{ma_raman_2018}. Second, the signs of diagonal elements of the $M$ and $Q$ tensors are not determined since these changes in the relative signs of diagonal elements will only modify the relative orientations of the interaction tensors without altering the hyperfine NMR spectra \cite{cruzeiro_characterization_2018,Longdell_quantum_2003}. The signs of diagonal elements of the $Q$ tensor in the ground state can be determined by the ordering of hyperfine levels at zero field, while the signs of the $M$ tensor can be determined by measuring the branching ratio of optical transitions.

The final fitting results are presented in Fig. \ref{fig:spectral} and Tab. \ref{tab:fit}. The fitting error is reduced to approximately 9 kHz per peak, which is close to the experimental error. The errors shown in Tab. \ref{tab:fit} are estimated using the non-linearly fitted covariance matrix in the laboratory coordinate system. The uncertainties are exclusive of the calibration of the coils which is expected to be less than 2\%. In addition, directions of D$_1$ and D$_2$ axes are listed in Tab. \ref{tab:fit}. Note that the measurement of the D$_1$ axis is not precise due to non-strictly horizontal or vertical axes defined by the XYZ coils. The error of the angles relative to the crystal axis system is estimated to be 5$^{\circ}$ because of manual placement. The polarization of incident light with the strongest absorption of the crystal is measured, which is close to the horizontal. The horizontal direction (x axis) is projected onto the D$_1$-D$_2$ plane, which is perpendicular to the C$_2$ axis, as a reasonable approximation for the D$_1$ axis \cite{ma_raman_2018}. As a result, the D$_1$ axis is estimated to be in the direction of [0.9993,0.0004,-0.0364] in the XYZ frame, and the D$_2$ axis is perpendicular to the D$_1$ and C$_2$ axes. We further provide M$_1$ and Q$_1$ tensors for the ground ($^3$H$_4$) and excited ($^1$D$_2$) state of Pr$^{3+}$:Y$_2$SiO$_5$ in the (D$_1$, D$_2$, b) basis as follows,
\begin{equation}
\begin{split}
Q_1^g &= \left[
\begin{matrix}
-0.234&0.471&-0.136\\
0.471&-0.996&-0.455\\
-0.136&-0.455&-0.099
\end{matrix}
\right]MHz,\\
M_1^g &= \left[
\begin{matrix}
3.232&-0.527&-0.188\\
-0.527&4.390&-0.457\\
-0.188&-0.457&2.288
\end{matrix}
\right]kHz/G,\\
Q_1^e &= \left[
\begin{matrix}
-0.278&-0.336&0.076\\
-0.336&-0.079&0.235\\
0.076&0.235&-0.295
\end{matrix}
\right]MHz,\\
M_1^e &= \left[
\begin{matrix}
1.561&-0.013&0.114\\
-0.013&1.524&0.150\\
0.114&0.150&1.555
\end{matrix}
\right]kHz/G.
\end{split}
\end{equation}

\begin{table}[tb]
\renewcommand{\arraystretch}{1.2}
\setlength{\abovecaptionskip}{0pt}  
\setlength{\belowcaptionskip}{10pt}
\caption{\label{tab:fit}The fitting results of the effective Hamiltonian parameters for the ground ($^3$H$_4$) and excited ($^1$D$_2$) state of Pr$^{3+}$:Y$_2$SiO$_5$ in the laboratory frame. The uncertainty in the calibration of the coils is not contained.}
\centering
\begin{tabular}{ccccc}
\toprule
Quantity & $^3$H$_4$ & Uncertainty & $^1$D$_2$ & Uncertainty\\
\midrule
    $\alpha_M$(degrees) & 16.0761 & 0.0048 & 142.73 & 0.12 \\
    $\beta_M$(degrees) & -72.572 & 0.042 & 87.00 & 0.21 \\
    $\gamma_M$(degrees) & -80.549 & 0.020 & -139.07 & 0.11 \\
    $g_1$(kHz/G) & 4.6459 & 0.0039 & 1.732704 & 0.000060 \\
    $g_2$(kHz/G) & 2.1125 & 0.0011 & 1.351342 & 0.000050 \\
    $g_3$(kHz/G) & 3.1497 & 0.0020 & 1.559799 & 0.000028 \\
    $\alpha_Q$(degrees) & 111.94258 & 0.00047 & 42.2215 & 0.0095 \\
    $\beta_Q$(degrees) & 73.2580 & 0.0076 & -60.7753 & 0.0072 \\
    $\gamma_Q$(degrees) & -45.794 & 0.057 & -19.1996 & 0.0016 \\
    $E$(MHz) & -0.305891 & 0.000076 & 0.216943 & 0.000020 \\
    $D$(MHz) & -1.32776 & 0.00034 & -0.65140 & 0.00032 \\
    $\theta_{C_2}$(degrees)& 2.169 & 0.092 \\
    $\phi_{C_2}$(degrees)& 15.99 & 0.42 \\
    $\theta_{D_1}$(degrees)& 92.0859 \\
    $\phi_{D_1}$(degrees)& -0.0218 \\
    $\theta_{D_2}$(degrees)& 90.5980 \\
    $\phi_{D_2}$(degrees)& 90.0000 \\
\bottomrule
\end{tabular}
\end{table}

\section{\label{spin}Spin coherence measurement}
After obtaining the spin Hamiltonians, we pay attention to the spin coherence time of Pr$^{3+}$ at site 2. The spin coherence time plays a prominent role in implementations of quantum information technology. Spin echo with Raman heterodyne detection is employed to obtain the spin coherence time $T_2$ \cite{hahn_spin_1950}.

First, a $\pi$/2 pulse drives the ions into a superposition state. After free evolution for time $\tau$, the inhomogeneous dephasing is removed by applying a $\pi$ pulse to refocus the components of the Bloch vector \cite{beavan_demonstration_2009,heinze_coherence_2014}. After another time $\tau$, a spin echo appears due to all spin coherence rephase again. An optical probe pulse is applied during the spin echo and the signal is captured by a photodetector and then observed on the oscilloscope after demodulation. In the experiment, the RF pulses are generated by the arbitrary waveform generator and amplified by the power amplifier. The duration and power of the $\pi$ pulse optimized to obtain the strongest signal of the spin echo are 10 $\mu$s and 300 W in our experiment. The amplitude of the spin echo decreases with the evolution time $\tau$. Thus, we can determine the coherence lifetime $T_2$ by the intensity as a function of the total evolution time 2$\tau$ \cite{fraval_method_2004}:
\begin{equation}
I(2\tau) = I_0\exp[-(\frac{2\tau}{T_2})^n],\label{eq:coherence fiting}
\end{equation}
where $n$ is any number between 1 and 2 depending on the spectral diffusion process and timescale. 

\begin{figure}[tb]
\includegraphics[width=\textwidth]{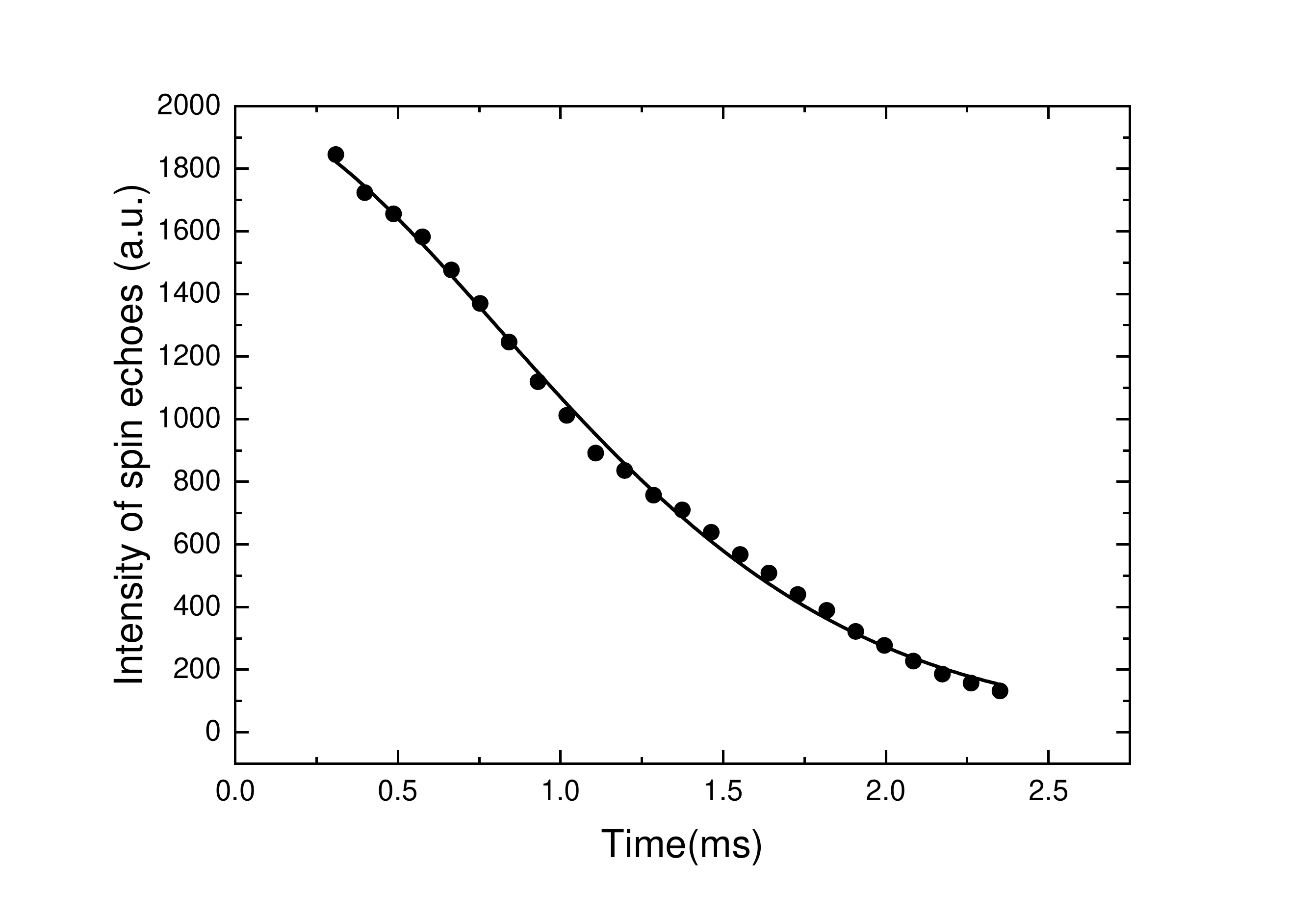}
\caption{\label{fig:zero}The decay of the spin echo energy vs coherence decay time $\tau$ at zero magnetic fields. The solid line is fitted based on the Eq. \ref{eq:coherence fiting} ($n$ = 1.73) and the spin coherence lifetime is  $T_2$ = 2.6$\pm$0.1 ms. The offset of the fit curve is 27$\pm$24.}
\end{figure}

Fig. \ref{fig:zero} shows the decay of the spin-echo signal in zero magnetic field.  Based on the Eq. \ref{eq:coherence fiting}, the measured data are fitted and the spin coherence lifetime ($T_2$) is 2.6$\pm$0.1 ms. This value is more than five times longer than that measured for site-1 Pr$^{3+}$ in Y$_2$SiO$_5$ \cite{ham_frequency-selective_1997}.

\section{\label{ZEFOZ}Searching the ZEFOZ magnetic fields}

\begin{figure}[tb]
\includegraphics[width=\textwidth]{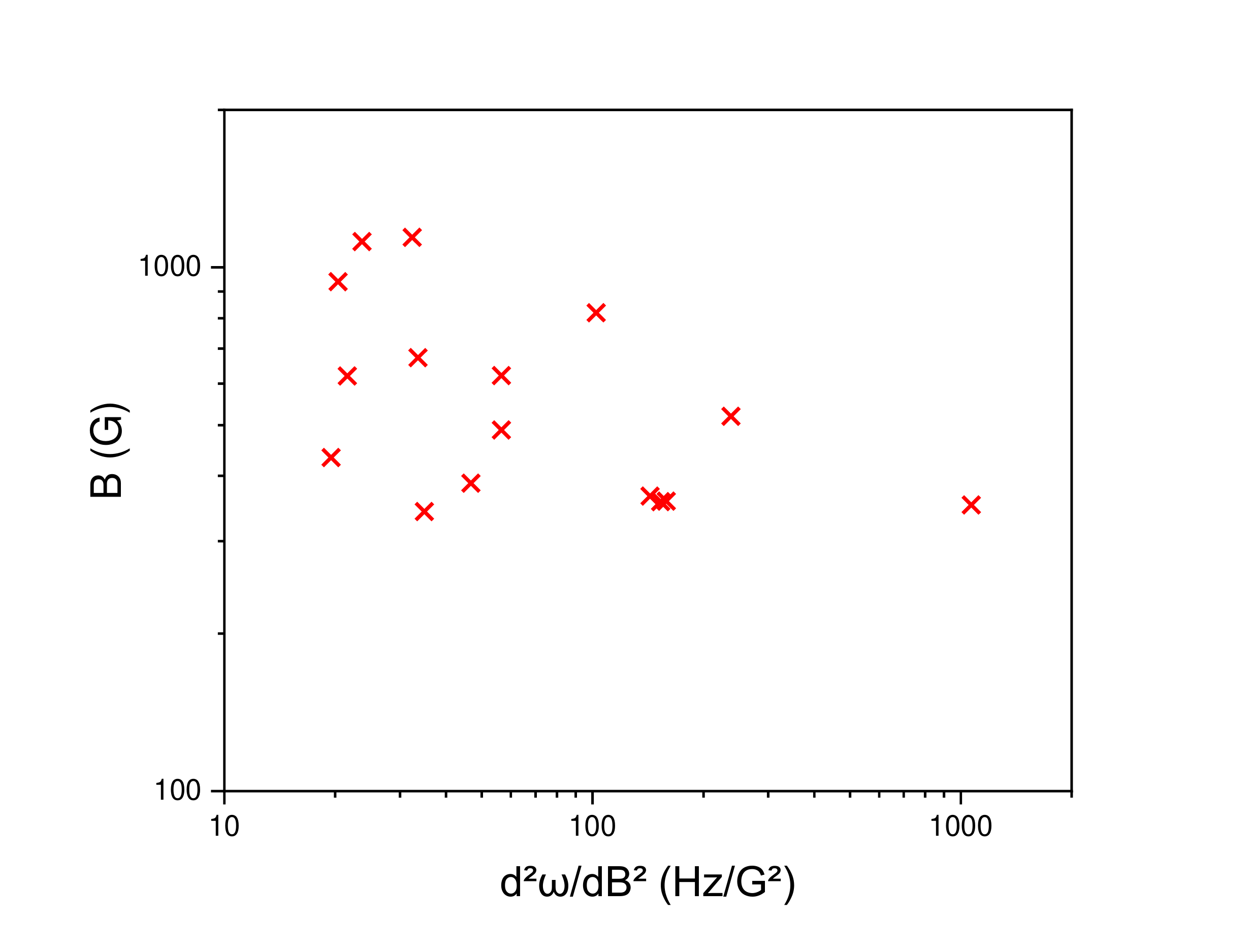}
\caption{\label{fig:level_ZEFOZ_fined}All ZEFOZ magnetic fields are identified for Pr$^{3+}$ at site 2 in Y$_2$SiO$_5$. The scatter plot is magnetic fields vs the second order Zeeman coefficient. The smallest second order Zeeman coefficient in the figure occurs at $[255.60,80.55,341.05]$G with a hyperfine transition frequency of 3.13 MHz.}
\end{figure}

The obtained Hamiltonians pave the way to search the zero first order Zeeman magnetic fields for Pr$^{3+}$ at site 2 in Y$_2$SiO$_5$. A ZEFOZ magnetic field refers to a certain field \cite{fraval_method_2004}, in which transition frequency under consideration is no more linearly dependent on the magnetic field. As a result, this leads to a weaker interaction between the praseodymium ions and the spins in the host lattice, which can significantly prolong the spin coherence time ($T_2$) \cite{fraval_method_2004,heinze_coherence_2014,fraval_pry_2004,lovric_hyperfine_2011}. Fig. \ref{fig:level_ZEFOZ_fined} shows all ZEFOZ magnetic fields found for Pr$^{3+}$ at site 2 in Y$_2$SiO$_5$. The method of finding the ZEFOZ magnetic fields has been detailed in a previous work \cite{longdell_characterization_2006}. The Zeeman gradient vector and curvature tensor are calculated using first and second order time independent perturbation theory. The ZEFOZ transition where the Zeeman gradient vector is zero is found by using the iterative algorithm in the work. All ZEFOZ transitions are given by searching on a three dimensional grid of magnetic field values.

The spin coherence time is related to the first and second order Zeeman coefficients using a simple model as \cite{zhong_optically_2015}
\begin{equation}
    \frac{1}{\pi T_2} = S_1\cdot\Delta B+\Delta B\cdot S_2\cdot\Delta B,\label{eq:projected_T2}
\end{equation}
where $T_2$ is the coherence time, $S_1$ and $S_2$ represent the first and second order Zeeman coefficients respectively, and $\Delta B$ is the degree of field fluctuation. $\Delta B$ was set at 0.08 G according to the previous publication \cite{zhong_optically_2015}. According to Eq. \ref{eq:projected_T2}, we estimate the spin coherence time for all ZEFOZ fields according to the $S_1$ and $S_2$ parameters calculated from the Hamiltonian. The best ZEFOZ transition is the 3.13 MHz transition at the field of $[255.60,80.55,341.05]$G in the lab frame, with a $S_1$ of 0.36 Hz/G and a $S_2$ of 19.5 Hz/G$^2$. The calculated spin coherence time is 2.1 s, which is more than two times longer than the 0.82 s expected for site-1 Pr$^{3+}$ in Y$_2$SiO$_5$ with a $S_2$ of 60 Hz/G$^2$ \cite{longdell_characterization_2006} in a perfect aligned ZEFOZ field. The ZEFOZ field can enable long-term spin-wave optical storage combined with various quantum memory protocols \cite{de_riedmatten_solid-state_2008,jobez_coherent_2015,chaneliere_storage_2005,Moiseev-complete-2001,Kutluer-Spectral-2016}.

\section{\label{con}Conclusion}
The hyperfine interactions of the ground state ($^3$H$_4$) and excited state ($^1$D$_2$) of Pr$^{3+}$ at site 2 in Y$_2$SiO$_5$ have been characterized by using Raman heterodyne detected NMR. Comparing to Pr$^{3+}$ at site 1, it is noteworthy that the hyperfine transition has an extended coherence time for Pr$^{3+}$ at site 2. An optimized ZEFOZ field is identified according to the predictions from the hyperfine Hamiltonian.

\section*{Acknowledgement}
This work was supported by the National Key R\&D Program of China (No. 2017YFA0304100), National Natural Science Foundation of China (Nos. 11774331, 11774335, 11504362, 11821404 and 11654002), Anhui Initiative in Quantum Information Technologies (No. AHY020100), Key Research Program of Frontier Sciences, CAS (No. QYZDY-SSW-SLH003), Science Foundation of the CAS (No. ZDRW-XH-2019-1) and Fundamental Research Funds for the Central Universities (No. WK2470000026, No. WK2470000029).


\end{document}